\documentclass[preprint,aps,prd,nofootinbib,superscriptaddress]{revtex4}

\usepackage[T1]{fontenc}
\usepackage[latin9]{inputenc}
\usepackage{amsmath}
\usepackage{graphicx}
\usepackage{tikz-feynman}
\usepackage{simplewick}
\usepackage{slashed}
\usepackage{subfigure}

\begin{document}

\preprint{\vbox{\hbox{WSU-HEP-1908, INT-PUB-19-062}}}

\title{\boldmath Semileptonic decays of heavy mesons with artificial neural networks}

\author{Cody M. Grant}
\affiliation{Department of Physics and Astronomy\\
        Wayne State University, Detroit, MI 48201, USA}

\author{Ayesh Gunawardana}
\affiliation{Department of Physics and Astronomy\\
        Wayne State University, Detroit, MI 48201, USA}
      
\author{Alexey A.\ Petrov}
\affiliation{Department of Physics and Astronomy\\
        Wayne State University, Detroit, MI 48201, USA}
\affiliation{Leinweber Center for Theoretical Physics\\
        University of Michigan, Ann Arbor, MI 48196, USA}

\date{\today}

\begin{abstract}
Experimental checks of the second row unitarity of the Cabibbo-Kobayashi-Maskawa (CKM) matrix involve extractions
of the matrix element $V_{cd}$, which may be obtained from semileptonic decay rates of $D$ to $\pi$. 
These decay rates are proportional to hadronic form factors which parameterize how the quark $c \to d$ transition 
is realized in $D \to \pi$ meson decays. The form factors can not yet be analytically computed over the whole range of available 
momentum transfer $q^2$, but can be parameterized with a varying degree of model dependency. We propose using artificial 
neural networks trained from experimental pseudo-data to predict the shape of these form factors with a prescribed uncertainty. 
We comment on the parameters of several commonly-used model parameterizations of semileptonic form factors.
We extract shape parameters and use unitarity to bound the form factor at a given $q^2$, which then allows us to bound the CKM 
matrix element $|V_{cd}|$. 
\end{abstract}
\maketitle

\section{Introduction}
Studies of exclusive semileptonic decays of heavy mesons play an important role in understanding the dynamics of the strong interaction.
They may also provide additional constraints on physics beyond the standard model (SM) \cite{Artuso:2008vf}. Such searches,
recently performed in $B$ decays yielded tantalizing results in measurements related by lepton universality requirements, i.e. 
by the requirement that couplings of leptons to gauge bosons be independent of the lepton flavor. It is interesting to see if similar anomalies 
exist in semileptonic decays of charmed particles if higher precision data are available \cite{Ablikim:2018evp,Ablikim:2018frk,Yuan:2019zfo,Riggio:2017zwh}.

Accurate theoretical description of such transitions is also needed for the extraction of relevant Cabbibo-Kobayashi-Maskawa (CKM) matrix elements. 
In particular, decays of charmed $D^0$, $D^+$, or $D_s$ mesons provide the simplest way to determine the magnitudes of quark mixing parameters 
$V_{cs}$ or $V_{cd}$ \cite{Riggio:2017zwh,Amhis:2019ckw}. 
Extractions of these CKM matrix elements from experimentally measured semileptonic decay rates are done with the knowledge of matrix elements 
of quark currents that describe strong interaction effects. This implies that accurate description of semileptonic transitions is also needed for improvement of our 
understanding of quark hadronization mechanisms in Quantum Chromodynamics (QCD). A hadronic transition between two mesons in exclusive semileptonic decays 
makes it a suitable system to theoretically analyze matrix elements of flavor changing currents, which are usually parameterized in terms of 
momentum-dependent form factors. In semileptonic decays of charmed mesons, the form factors that describe the hadronic
part of the decay amplitudes are conventionally introduced as 
\begin{equation}\label{DPseudoscalar}
\langle K(\pi) (p_{K(\pi)}) | \bar q \gamma_\mu c | D (p_D) \rangle = F_+(q^2) \left(P_\mu - \frac{m_D^2-m_{K(\pi)}^2}{q^2} q_\mu \right) +
 F_0(q^2) \frac{m_D^2-m_{K(\pi)}^2}{q^2} q_\mu \ ,
\end{equation}
where $P = p_D + p_{K(\pi)}$ and $q = p_D - p_{K(\pi)}$. Experimental studies of these form factors are performed through the analysis of the differential decay rate
 $d\Gamma/dq^2$. In the simplest cases where the mass of the final state lepton can be neglected, the differential decay rate can be written as 
\begin{equation}\label{DifDistr}
\frac{d\Gamma(D \to K(\pi) \ell \nu_\ell)}{dq^2} =
\frac{G^2_F \left|V_{cq}\right|^2}{24 \pi^3} \left|{\bf p}_{K(\pi)}\right|^3 \left|F_+(q^2)\right|^2, 
\end{equation}
where $\left|{\bf p}_{K(\pi)}\right|$ is the magnitude of the $K(\pi)$ 3-momentum vector in the $D$-meson rest frame. As can be seen from Eq.~(\ref{DifDistr}), only a single form factor, 
$F_+(q^2)$, contributes. 

Accurate calculations of the non-perturbative form factors $F_{+/0}(q^2)$ in the whole momentum range are very challenging. Aside from lattice QCD \cite{Aoki:2019cca} and/or 
QCD sum rule (QCDSR) \cite{Khodjamirian:2009ys} calculations of matrix elements of hadronic currents in exclusive decays, we are currently lacking a complete non-perturbative description 
of hadronic form factors. While both lattice QCD and QCDSR computations of form factors are improving, at the moment they only provide model-independent predictions for
$F_+(q^2)$ at limited regions of $q^2$. 

Rather general arguments based on analyticity of $F_+(q^2)$ have been used to place general constraints on the shapes of the form factors. 
A popular approach that rigorously employs analyticity requirement involves the so-called $z$-expansion, where a series expansion of the form factor around some point 
$t=q^2$ is improved by making a conformal transformation to the parameter $z$ \cite{Boyd:1994tt}, 
\begin{equation}\label{Zexp}
z(q^2)=\frac{\sqrt{t_+-t_0}-\sqrt{t_+-q^2}}{\sqrt{t_+-t_0}+\sqrt{t_+-q^2}},
\end{equation}
which maps the interval $-\infty < q^2 < t_+$ onto the line segment $-1<z<1$. Here $t_0$ is a free parameter that corresponds to the values of $q^2$ that maps 
onto $z=0$, and $t_{\pm}= (m_D\pm m_\pi)^2$. The form factor can be expanded in $z$ as
\begin{equation}\label{FF_z}
F_+(q^2)  =  \frac{1}{\Phi(q^2, t_0)} \sum_{k=0}^\infty a_k(t_0) z^k(q^2,t_0)
\end{equation}
where  $\Phi(q^2, t_0)$ is an arbitrary function that is analytic anywhere but the unitarity cut \cite{Boyd:1994tt,Becher:2005bg}. Note that $\Phi(q^2, t_0)$ is often written as 
$\Phi(q^2, t_0) = P(q^2) \phi(q^2, t_0)$, with the Blaschke factor $P(q^2) = z(q^2, m_V^2)$ if there are poles present in between $q^2=0$ and the beginning of the unitarity cut, 
as in $B\to \pi$ transitions where $m_V=m_{B^*}$ \cite{Ananthanarayan:2011uc,Grinstein:2015wqa}. Note that $P(q^2) = 1$ for the $D \to \pi \ell \nu$ transition.
The expansion in Eq.~(\ref{FF_z}) is converging rapidly, so only
a few terms in the expansion are really needed\footnote{See however \cite{DescotesGenon:2008hh} for a discussion of possible shortcomings of this approach.}. 
Lattice QCD \cite{Aoki:2019cca} or QCD sum rule \cite{Khodjamirian:2009ys} results can be used to constrain the coefficients 
$a_k$ to provide a model-independent parameterization of the form factor.

As it stands, phenomenological parameterizations of the form factors are also often used \cite{Faustov:2019mqr}. The most common parametrization is a 
``single pole'' shape, where the pole refers to the lowest mass vector resonance formed in the t-channel with quantum numbers of the quark current. 
For example, in the decay $D\to \pi e \bar\nu_e$ the dominant pole is the $D^\star$, a vector state with $1^{-}$ quantum numbers,
\begin{equation}\label{FF_Pole}
F_+^{\rm pole}(q^2) = \frac{F_+(0)}{1-{\hat q}^2},
\end{equation}
where $F_+(0)$ is the value of the form factor at zero momentum recoil that has to be fixed either from the lattice QCD or from other arguments,
and $\hat{q}^2=q^2/m_{D^*}^2$. While physical masses of the states $D^*(2010)$ (for $D\to \pi$ transition) or $D^*_s(2112)$ (for $D\to K$ transition) could be used, 
the mass $m_{D*}$ is often taken as a fit parameter, as there is no reason to believe that the lowest-lying pole would saturate the form factor over the whole available 
kinematical range. More complicated shapes, with more effective poles, are also available \cite{Artuso:2008vf}, 
\begin{equation}\label{FF_ManyPoles}
F_+(q^2)=\frac{F_+(0)}{(1-\alpha)}\frac{1}{1-q^2/m_V^2}+
\sum_{k=1}^N \frac{\rho _k}{1-\frac{1}{\gamma_k}\frac{q^2}{m_V^2}},
\end{equation}
where $\alpha$  determines the strength of the dominant pole, $\rho_k$ gives the strength of the $k$th term in the expansion, and
$\gamma _k=m_{V_k}^2/m_V^2$, with $m_{V_k}$ representing masses of the higher mass states with vector quantum numbers. In principle, a form factor can be 
approximated to any desired accuracy by introducing a large number of effective poles. Keeping the number of terms in this expansion manageable, a popular 
parameterization due to Becirevic and Kaidalov (BK) \cite{Becirevic:1999kt} is often used, representing the $N=1$ truncation of the expansion in Eq.~(\ref{FF_ManyPoles}),  
\begin{equation}\label{FF_BK}
F_+^{BK}(q^2)  =  \frac{F_+(0)}{(1-\hat{q}^2)(1-a_{BK}\hat{q}^2)},
\end{equation}
where $a_{BK}$ is a fit parameter. As with the case of a single pole shape in Eq.~(\ref{FF_Pole}), a good fit to experimental distribution can be obtained 
if $m_V$ is regarded as a fit parameter as well. A further extension of the BK parameterization was proposed by Ball and Zwicky (BZ) \cite{Ball:2004ye,Su:2010my},
\begin{equation}\label{FF_BZ}
F_+^{BZ} (q^2) = \frac{F_+(0)}{1-\hat{q}^2} \left(1+\frac{r_{BZ}\hat{q}^2}{1-a_{BZ}\hat{q}^2}\right),
\end{equation}
where $r_{BZ}$ and $a_{BZ}$ are the shape parameters. Note that the parameterization of the form factor in BZ model employed in Eq.~(\ref{FF_BZ}) \cite{Su:2010my}
can be related to the original BZ-parameterization \cite{Ball:2004ye}
\begin{equation}\label{FF_BZOrig}
F_+^{BZ} (q^2) = \frac{r_1}{1-\hat{q}^2}  + \frac{r_2}{1-a_{BZ}\hat{q}^2},
\end{equation}
by the identification $F_+ (0) = r_1+r_2$ and $r_{BZ} = \left(a_{BZ}-1\right) r_2/(r_1+r_2)$.
Note that $a_{BZ}$ represents parameterization of the continuum states above $D^*$ and therefore $a_{BZ}<1$. 

All form factor parameterization discussed above represent physically-motivated ways to describe hadronic input. Yet, a question might be asked then what uncertainty 
should be assigned to the {\it choice} of a particular shape of the fit function. In other words, we will be interested if choosing a specific functional form for the form factor 
induces a bias in the interpretation of results of an experimental analysis.

This question may be addressed in the framework of machine learning (ML) approach, in particular, it can be investigated with the help of artificial neural networks (ANN). 
Based on the Kolmogorov-Arnold representation theorem \cite{Kolmogorov57},
it has been shown that ANN can be used as an unbiased estimator of data \cite{NNEstimator,NNEstimator2}. This fact has been used by the NNPDF collaboration to 
parameterize nucleon's parton distribution functions \cite{Forte:2002fg,Ball:2014uwa,Rojo:2006nn}, and in form factor analysis of nucleon data \cite{Graczyk:2010gw,Alvarez-Ruso:2018rdx}. 
In this paper we shall build a statistical interpolating model based on ANNs that contains information on experimental uncertainties and correlations, but does not introduce 
theoretical bias. Following  \cite{Forte:2002fg,Ball:2014uwa}, we employ an approach based on multilayer feed-forward neural networks trained using the back-propagation learning
algorithm.

\section{Neural Networks}
\subsection{Basic facts about neural networks}

With the recent explosion of interest in machine learning, artificial neural networks are now widely employed in analyses in experimental particle physics. Their use in jet-finding
algorithms and other applications are well known \cite{Carleo:2019ptp}. Roughy speaking, a neural network is represented by a certain non-linear function that connects input and output data.
This leads to another feature of ANNs which we explore in this paper: their ability to provide unbiased universal approximants to incomplete data \cite{NNEstimator,NNEstimator2}. 
\begin{center}
\begin{figure}[h]
\center
\includegraphics[scale=0.9]{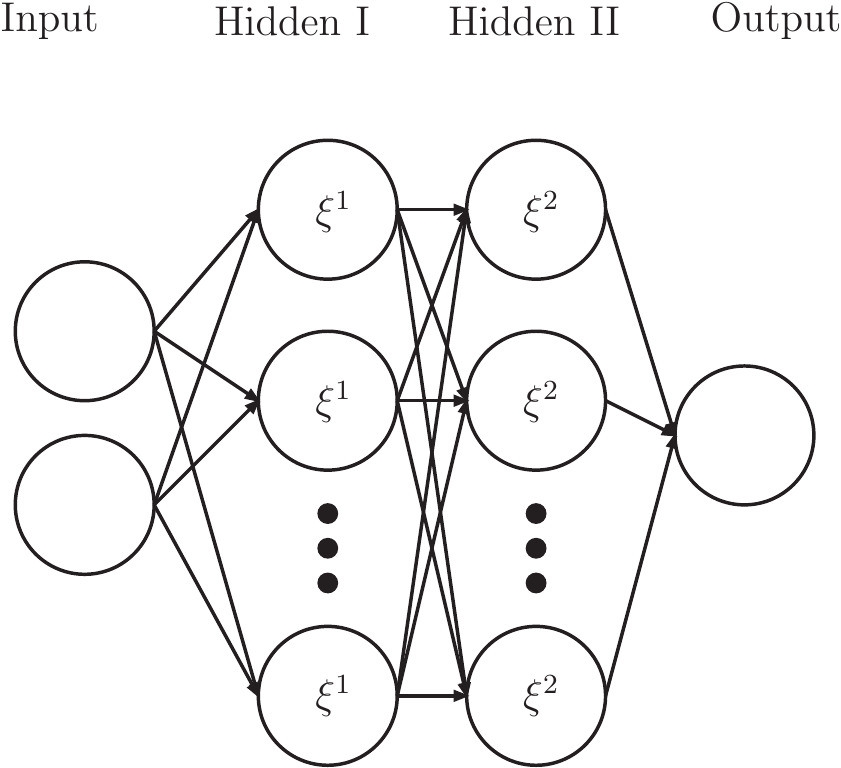}
\caption{A sample structure of an artificial neural network with two hidden layers. \label{Fig:ANN}}
\end{figure}
\end{center}
An ANN is built to mimic the structure of human neurons and consists of a set of interconnected units (see Fig.~\ref{Fig:ANN}) called {\it neurons} or {\it nodes}. The activation state of a neuron is determined 
as a function of the activation states of the $i$ neurons connected to it. Each pair of neurons is connected by a synapsis, characterized by a weight, which we call $\omega_i$. We also introduce a 
set of $\theta_i$, representing thresholds for each neuron to ``fire". Each ANN contains several groups of neurons called {\it layers}. The first layer is called an input layer. It provides input 
information that is to be approximated. In this paper the input information is the value of $q^2$ for each bin in $q^2$ distribution of the CKM matrix element times the semileptonic form factor. 
We find it convenient in this work to work 
with an input layer that contains two nodes, as we shall explain later. The final layer is the output layer. It gives the value of form factor for each $q^2$ along with its uncertainty. Layers between the input 
and output are conventionally called {\it hidden}. In this work we employ ANN with two hidden layers of 100 nodes each. The ANN is {\it trained} when optimal sets of weights and thresholds are 
determined such that ANN reproduces the training data within a given uncertainty. This is achieved by minimizing the error function,
\begin{eqnarray}\label{error_fun}
E[{\omega,\theta}]\equiv\frac{1}{2}\sum_{A=1}^{n_p}
(o(q^2_{A})-y_{A})^2\,,
\end{eqnarray}
where $n_p$ is the number of pseudo-data used to train an ANN, $o( q^2_{A})$ is the output, which is given by the ANN's fit for a given input data $q^2_{A}$.
The target data point for our paper, $y_A$, is obtained from the magnitude of the CKM matrix element times the semileptonic form factor, $\left|V_{cd} F_+(q^2)\right|$.
The differential distribution of Eq.~(\ref{DifDistr}) is proportional to its square.

The $o(q^2_{A})$ is obtained using forward propagation. In order to achieve this we pass the input through a network of hidden nodes. The output from the first hidden layer with 
$n_1$ number of nodes is
\begin{eqnarray}\label{first_layer_output}
\xi^{[1]}=g\left(\sum_{i=1}^{n_1}\omega_{i}^{[1]}q^2-\theta^{[1]}\right).
\end{eqnarray}
In this equation the response of each neuron is given by 
\begin{eqnarray}\label{sigmoid}
g(x)\equiv\frac{1}{1+e^{-x}},
\end{eqnarray}
which is the \textit{sigmoid} activation function, and the summation over the $q^2$ data points is implied.

The $\xi^{[1]}$ is then used as an input for the second hidden layer with $n_2$ number of hidden nodes, and so on. The process is continued until the output layer of ANN is reached. In general, 
we can construct the output from $\ell$th hidden layer with $n_\ell$ number of nodes as 
\begin{eqnarray}\label{general_output}
\xi^{[\ell]}=g\left(\sum_{i=1}^{n_{\ell}}\omega_{i}^{[\ell]}\xi^{[\ell-1]}- \theta^{[\ell]} \right)  .
\end{eqnarray}
where $\xi^{[\ell-1]}$ is the output from the $(\ell-1)$th layer. The fit of the $L$ layer ANN $o(q^2)$ is then defined as   
\begin{eqnarray}
o(q^2)=\xi^{[L]}.
\end{eqnarray}
In the training process the thresholds and weights need to be adjusted so the output represented the training data with a set precision,
so the error function in Eq.~(\ref{error_fun}) need to be minimized. It is common to use the method of steepest descent for this purpose. Instead, 
we decided to use the non-linear conjugate gradient (NLCG) method \cite{Nocedal:2000sp, CGmethod} to minimize Eq.~(\ref{error_fun}). 
In each iteration the $\omega_i$ and the $\theta_i$ update as
\begin{eqnarray}\label{steepest_descent}
\delta\omega^{[\ell]} &=& -\eta\frac{\partial
E}{\partial\omega^{[\ell]}}\,, \nonumber \\ \\ \nonumber
\delta\theta^{[\ell]} &=& - \eta\frac{\partial
E}{\partial\theta_{i}^{[\ell]}},
\end{eqnarray} 
where $\eta$ is the learning rate at a given iteration. The NLCG method employed here does not require a pre-defined learning rate. 
The learning rate is initially determined by using line search algorithms \cite{Nocedal:2000sp}, and then iteratively updated based on 
the gradients that are in a conjugate direction to original gradient used in the line search algorithm. As it turns out, the NLCG method 
converges much faster than steepest descent method for the fits employed in this paper. For more details on the NLCG method, see 
Ref. \cite{CGmethod}. The gradients of the error function are obtained by using the method of back propagation \cite{Demuth:2014}. 
Back propagation can be thought of as a consecutive application of the chain rule. By applying the chain rule to the $L$th layer we find 
\begin{eqnarray}\label{back_prop_final_layer}
\Delta^{[L]}= g'(h^{[L]})[o(q^2)-y]
\end{eqnarray} 
where $g'(h^{[L]})$ is the derivative of the activation function with respect to $h^{[L]}$ and 
\begin{eqnarray}\label{h_of_L}
h^{[L]}=\sum_{i=1}^{n_{L-1}}\omega^{[L]}\xi_{i}^{[L-1]}-\theta^{[L]}
\end{eqnarray} 
The derivatives with respect to $\omega_i$ and $\theta_i$ for layer $L$ are given by
\begin{eqnarray}\label{delta}
 \frac{\partial
E}{\partial\omega_{i}^{[L]}} &=&
\Delta^{[L]}\xi_i^{[L-1]};\qquad
i=1,\ldots,n_{L-1}\,, \nonumber \\ \frac{\partial
E}{\partial\theta_{i}^{[L]}} &=& -
\Delta^{[L]}.
\end{eqnarray}
The output of Eq.~(\ref{back_prop_final_layer}) is used to obtain the derivatives of the $(L-1)$th layer, $\Delta_j^{[L-1]}$,
\begin{eqnarray}
\Delta_j^{[L-1]}=g'\l(h^{[L-1]})
\Delta_i^{[L]}\omega^{[L]}\,.
\end{eqnarray}
The procedure is repeated for the hidden layers to find derivatives of error function with respect to $\omega_i$ and $\theta_i$ in each layer, 
\begin{eqnarray}\label{delta1}
 \frac{\partial
E}{\partial\omega_{ij}^{[\ell]}} &=&
\Delta_i^{[\ell]}\xi_j^{[\ell-1]};\qquad
i=1,\ldots,n_\ell,\quad j=1,\ldots,n_{\ell-1}\,, \nonumber \\ 
\frac{\partial E}{\partial\theta_{i}^{[\ell]}} &=& -
\Delta_i^{[\ell]},\qquad i=1,\ldots,n_\ell,
\end{eqnarray}
Using these we can obtain the numerical gradient of the error function and find the corrections to the weights and thresholds.

\subsection{Neural network training}
Training of ANNs described in the previous section must be performed either on real or artificial data (pseudo-data). The pseudo-data is generated using as much experimental information as possible. 
It can be constructed with uncorrelated data, correlated data, normalized data, or some combination of all three. In this work we elected to follow \cite{Rojo:2006nn} and generate pseudo-data 
from the BES III experimental data set of \cite{Ablikim:2015ixa} employing Monte Carlo techniques. We chose to select only this experimental data set and not to include earlier experimental data
because the BES III data set includes both uncorrelated and correlated statistical and systematic uncertainties, with correlation matrices available. It would indeed be interesting to apply our
methodology to upcoming Belle II data. The artificial data is generated as 
\begin{equation} \label{PD}
\left|V_{cd} F_{+}(q^2)\right|^{\rm{(art)},(k)}_i = \left|V_{cd} F_{+}(q^2)\right|^{\rm (exp)}_i + r_{t,i}^{(k)}\sigma_{t,i} + \sum_{j=1}^{N_{\rm sys}}r_{{\rm sys},j}^{(k)}\sigma_{{\rm sys},ji} 
+ \sum_{m=1}^{N_{\rm stat}}r_{{\rm stat},m}^{(k)}\sigma_{{\rm stat},mi}
\end{equation}
where $i=1,...,N_{\rm data}$ is the number of experimental data entries considered, which is equal to the number of $q^2$ bins. These entries are used to generate $k=1,...,N_{\rm rep}$ of Monte 
Carlo ``replicas.'' These replicas are generated following the recipe of \cite{Rojo:2006nn}. The first term on the right-hand side of Eq.~(\ref{PD}) is the central value from the experimental data 
point for a given $q^2$ bin. The data points in the replicas are created from it by using the remaining three terms on the right hand side of Eq.~(\ref{PD}), which provide variation in pseudo-data samples.
They represent experimental uncertainties (total uncorrelated, correlated systematic, and correlated statistical, respectively) obtained from the experimental data. Each ``uncertainty term'' 
is multiplied by a Gaussian random number $r_{t,i}^{(k)}$, $r_{{\rm sys},j}^{(k)}$, or $r_{{\rm stat},m}^{(k)}$ which have a mean of zero and a standard deviation of one. 
The total uncorrelated uncertainty, $\sigma_{t,i}$, is defined as the quadratic sum of the uncorrelated systematic and statistical uncertainties, 
$\tilde{\sigma}_{{\rm sys},ji}$ and $\tilde{\sigma}_{{\rm stat},mi}$, respectively,
\begin{equation}
\sigma_{t,i} = \sum_{j=1}^{N_{u,sys}}\tilde{\sigma}_{{\rm sys},ji} + \sum_{m=1}^{N_{u,sys}}\tilde{\sigma}_{{\rm stat},mi}
\end{equation}
The correlation matrix elements, $\mbox{corr}(j,i)$, found in Ref. \cite{Ablikim:2015ixa} are related to $\sigma_{{\rm sys},ji}$ and $\sigma_{{\rm stat},li}$ as
\begin{equation}
\sigma_{j,i} = \sqrt{\tilde{\sigma}_i \tilde{\sigma}_j \ \mbox{corr}(j,i)},
\end{equation}
where $\tilde{\sigma}_i$ is the uncorrelated uncertainty in the $i$-th bin of data. The $q^2$ values were randomly generated with a flat prior across the entire $q^2$ bin. 
Every value of $d\Gamma^{\rm (art)}/dq^2$ has a different $q^2$ input generated for it. 

We take the pseudo-data we have generated and divide it up into 100 batches (one batch per network). Each batch has an average $q^2$ and a standard deviation relating to the $q^2$ 
values, which are used to scale the each value of $q^2$ which we have generated. Using the scaled $q^2$ data as a secondary input is recommended to improve the stability and the 
performance of ANNs \cite{patro2015normalization}. In particular, data standardization is a popular data scaling choice, and it is defined as 
$\tilde{q}^2_{ i\rho} = \left(q^2_{i\rho} - \bar{q}^2_{\rho}\right)/\sigma_\rho$, where $\rho$ is the batch number and $i$ is a single $q^2$ value in the batch. With this transformed data, 
each of our ANNs has the structure (2, 100, 100, 1), as the two hidden layers, each with 100 nodes, provide the most efficient structure without compromising the performance or accuracy. 
With a higher number of nodes, the ANN's fit would be more accurate, but the training speed would also be reduced. This data transformation, along with the conjugate gradient method, 
provides the minimum of the error function at $100$ iterations. In contrast, steepest descent method with a constant learning rate provides a comparable result only at 20000 iterations. 

%
\section{Form factor parameterization with neural networks}
\begin{figure}[h]
\includegraphics[scale=0.9]{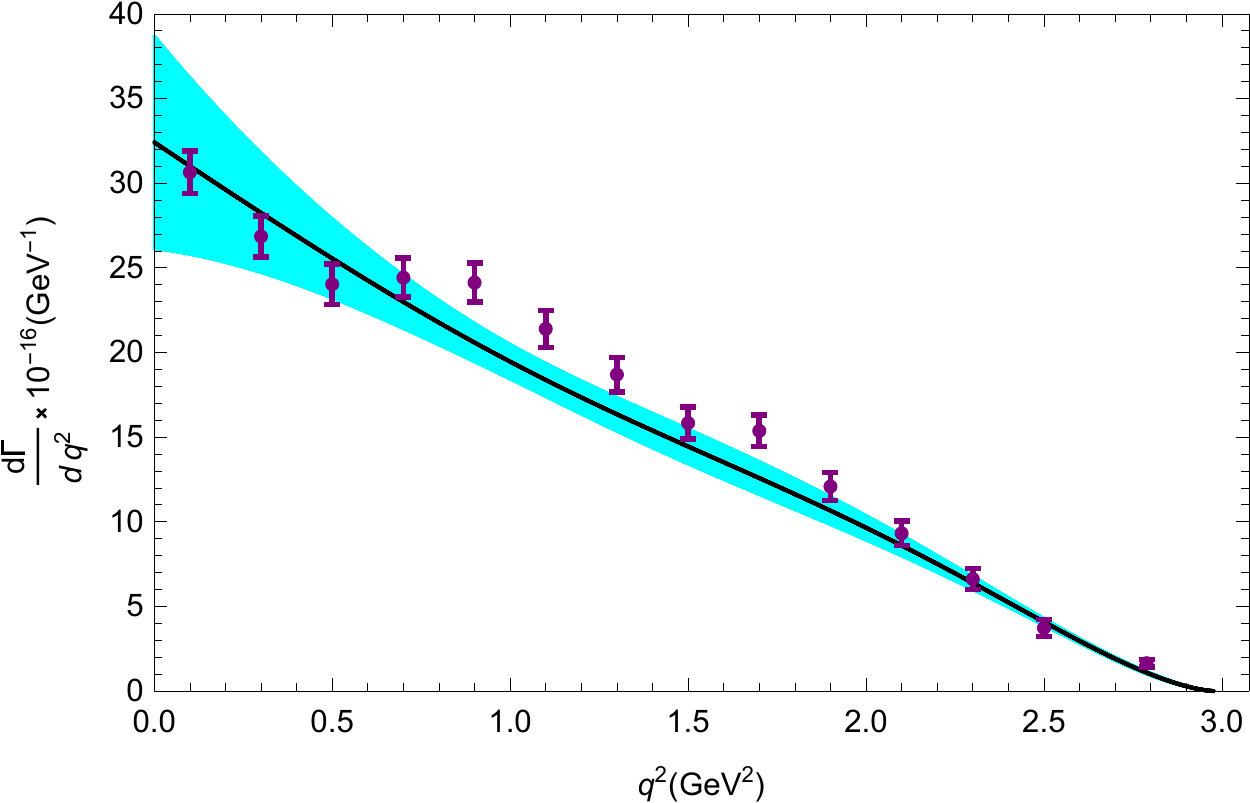}
\caption{Our averaged ANN result for the differential decay rate plotted against the experimental measurement. The purple data points are the experimental data from \cite{Ablikim:2015ixa}. 
The black and cyan curves are the average value and one standard deviation, respectively, from the output of our averaged ANN.}
\label{fig:diffGamma}
\end{figure}

\begin{figure}[h]
\includegraphics[scale=0.9]{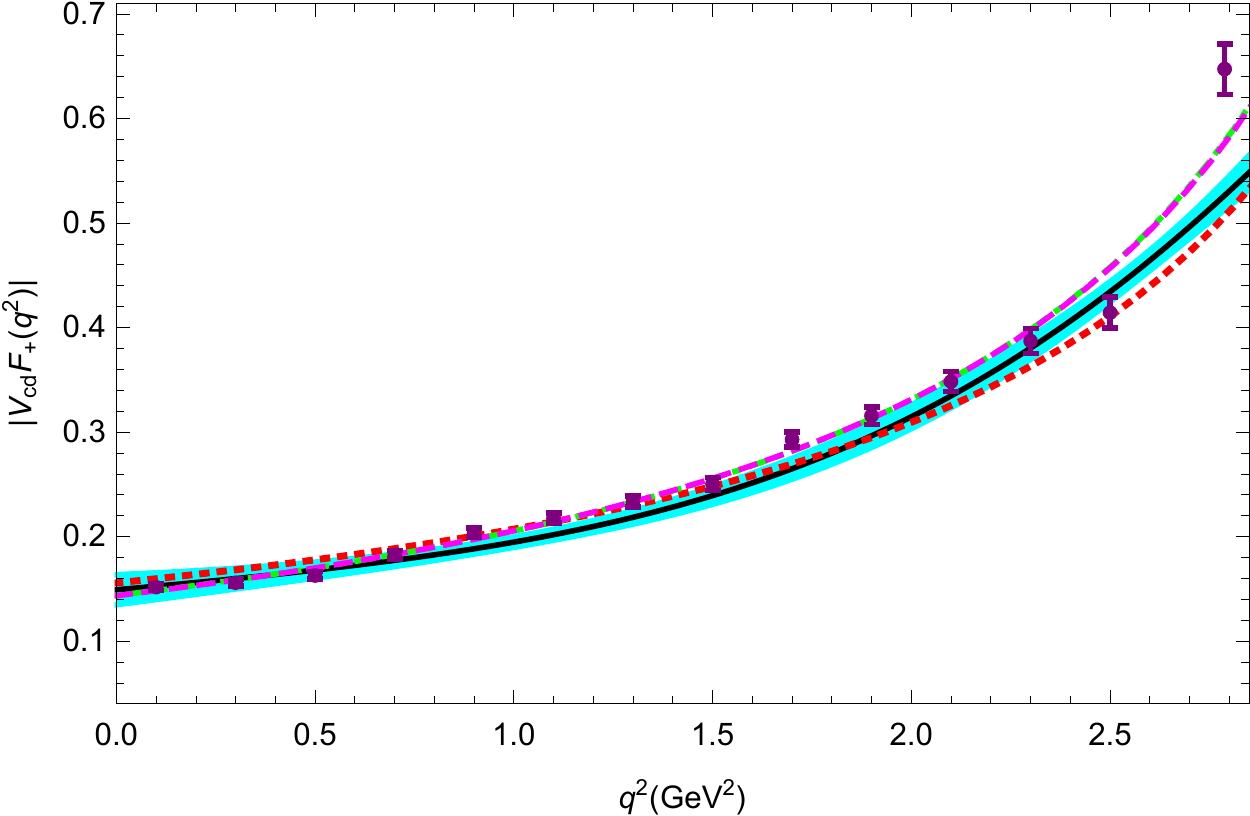}
\caption{ANN fits for $\left|V_{cd} F_+ (q^2)\right|$ plotted against the three models described in the text. The black and cyan curves are the average value and one 
standard deviation, respectively, from the output of our neural network. The dotted red curve is the simple pole model. The dot-dashed green curve is the modified 
pole model. The dashed magenta curve is the BZ model. The purple data points are calculated from the experimental data in Ref \cite{Ablikim:2015ixa}.}
\label{fig:VFwithModels}
\end{figure}

We generated $1.8\times10^{4}$ pseudo-data points for each $q^2$ bins for each network. 
After training all networks individually, we found the average ANN curve, with uncertainty, at every calculated $q^2$ value. The differential decay rate, 
$d\Gamma/dq^2$, and the $\left|V_{cd} F_{+}(q^2)\right|$ curves are shown in Fig. \ref{fig:diffGamma} and Fig.~\ref{fig:VFwithModels} respectively.
Further results of the ANN training and relevant graphs are available at the URL \url{s.wayne.edu/HEPMachineLearning}.

We would like to compare our results with some common form factor models: simple pole, the BK model (or modified pole), and the BZ model \cite{Becirevic:1999kt, Ball:2004ye, Su:2010my}. 
Since the ANN fits a product of $V_{cd}$ and $F_+(q^2)$, a direct comparison will be affected by the value of $V_{cd}$ that would have to be taken as an external parameter. 
With that in mind, we can compare $\left|V_{cd} F_+ (0)\right|$ obtained from the model fits and our ANN analysis of the semileptonic decay data.

A further insight into how well model-inspired parameterizations describe hadronic dynamics is possible if we expand the form factor around $q^2=0$, 
\begin{equation}\label{ff_q2_expansion}
\left|V_{cd} F_{+}\left(q^2\right)\right|=\left|V_{cd} F_{+}(0)\right|\left(1 + F_{1} q^2 + F_{2} q^4+...\right),
\end{equation}
and compare the coefficients $F_n$ of the higher order terms for the models to our averaged ANN output. We looked at the ratios of the $n$th derivative of the form factor divided 
by the form factor at $q^2=0$, 
\begin{equation}
F_n= \frac{1}{n! F_+(0)}  \left. \frac{d^n F_+(q^2)}{d(q^2)^n} \right|_{q^2=0} .
\end{equation}
We note that the first and second terms in Eq.~(\ref{ff_q2_expansion}), which are independent of the value of $V_{cd}$, are quite sensitive to the 
quark hadronization dynamics. In particular, drawing parallels to the discussion of the charge radius of the proton \cite{Paz:2019wfq}, the slope of $F_+(q^2)$ at $q^2=0$, 
denoted $F_1$,  encodes the information about the effective size of the volume where the quark transition takes place. We shall call the coefficient $F_1$ a transitional charge radius.

In order to compare $F_{1,2}$ of a particular model-inspired parameterizations to our ANN fits, we need to determine shape parameters for each form factor model. 
Other than the simple pole model, where we take the mass of the $D^*(2010)$ resonance as $m_{D^*}$, the parameters that need to be fit include $a_{BK}$ for the BK model, and
$a_{BZ}$ and $r_{BZ}$ for the BZ model. We obtain these shape parameters by fitting the model to the experimental data. Using this procedure we find 
$a_{BK} = 0.277 \pm 0.029$ for the BK model and $r_{BZ} = 0.295 \pm 0.123$ and $a_{BZ} = 0.192 \pm 0.555$ for the BZ model. 
We note that for each of these parameterizations the combination $\left|V_{cd} F_+ (0)\right|$ is also treated as a fit parameter. 

\begin{table}
 \begin{tabular}{|c|c|c|c|} 
 \hline\hline
 Form factor &\quad $\left|V_{cd} F_+(0)\right| \times 10^{-2}$ &\quad $F_{1} \times 10^{-1} \text{ GeV}^{-1}$ &\quad $F_{2} \times 10^{-1} \text{ GeV}^{-2}$ \\ [2ex] 
 \hline
 ~~ANN (this work)~~ &\quad $14.92 \pm 0.14$ &\quad $2.062 \pm 0.261$ &\quad $0.869 \pm 0.290$ \\ 
 \hline
 $F_+^{\rm pole} (q^2)$ &\quad $15.57 \pm 0.10$ &\quad $2.4830 \pm 0.0001$ &\quad $1.2330 \pm 0.0001$ \\ 
 \hline
$F_+^{\rm BK} (q^2)$ &\quad $14.37 \pm 0.16$ &\quad $3.170 \pm 0.072$ &\quad $1.669 \pm 0.055$ \\ 
 \hline
 $F_+^{\rm BZ} (q^2)$ &\quad $14.35 \pm 0.25$ &\quad $2.961 \pm 0.306$ &\quad $1.540 \pm 0.271$\\ 
 \hline \hline
\end{tabular} 
\caption{Form factor parameters at $q^2=0$ for this work and three common model choices. Note the unreasonably tiny uncertainty of the parameters $F_{1,2}$ of 
the pole form factor, which is related to the rigidity of the chosen functional form.}
\label{table:RatioTable}
\end{table}

The resulting values for $\left|V_{cd} F_+ (0)\right|$, $F_1$, and $F_2$ for the neural network parameterization and the model-inspired parameterizations can be found in 
Table \ref{table:RatioTable}. As we can see from the first column of Table \ref{table:RatioTable}, the values of $\left|V_{cd}F_+(0)\right|$ are consistent throughout the popular 
form factor models and are roughly consistent with our ANN study. The agreement is much worse for the parameters $F_{1,2}$: the ANN fits are consistently larger for the 
transitional charge radius $F_1$ and only marginally describing the $F_2$ parameter. It is likely that this happens due to rather rigid parameterizations of the model-inspired 
form factors, which artificially decrease possible uncertainties associated with them. This is particularly true for the simple pole parameterization $F_+^{\rm pole} (q^2)$: the 
uncertainty of $F_{1,2}$ is unreasonably small because once $\left|V_{cd} F_+ (0)\right|$ is fixed, the only uncertainties that can cause the spread in $F_{1,2}$ are the experimental 
uncertainties in the value of $m_{D^*}$, which are rather small. We conclude that it is possible that more effective poles need to be taken into account if model-inspired form 
factors are used for parameterizations of future experimental data. Graphically, the model fits compared to the artificial neural network fits and experimental data points are 
shown in Fig.~\ref{fig:VFwithModels}.

%

\section{Form factor bounds and their derivatives}
We can use our ANN fits to obtain separate bounds on the CKM matrix element $V_{cd}$ if we combine our fits with model-independent bounds on the hadronic
form factor imposed by analyticity and unitarity requirements \cite{Ananthanarayan:2011uc}. In order to do so and place an upper bound on $\left|F_{+}(0)\right|$, we would 
need to calculate moments of the heavy-light invariant amplitude $\Pi_{+}(q^2)$, which we denote by $\chi_{+}^{(n)}$. They are defined by the relation,
\begin{equation}\label{dispRel}
\chi_{+}^{(n)}=\frac{1}{\pi}\int_{t_+}^{\infty} dt \frac{\text{Im}\ \Pi_{+}(t+i\epsilon)}{t^{n+1}},
\end{equation}
where $n$ denotes a specific moment and $t_+=(m_D+m_\pi)^2$. These moments can be computed in QCD. In addition, an inequality for the imaginary part 
of $\Pi_{+}(q^2)$, which holds for $t>t_+$,  
\begin{equation}
\text{Im}\ \Pi_{+}(t+i\epsilon) \geq \frac{3}{2} \frac{1}{48\pi} \frac{[(t-t_+)(t-t_-)]^{3/2}}{t^3}\left|F_+(t)\right|^2
\end{equation}
can be found via the unitarity sum of the $D\pi$ state spectral function in the isospin limit \cite{Ananthanarayan:2011uc}. This result leads to an inequality with respect to the moment,
\begin{equation}\label{inEqB4map}
\chi_{+}^{(n)} \geq \frac{1}{\pi}\int_{t_+}^{\infty} dt \rho_+^{(n)}(t) \left|F_+(t)\right|^2
\end{equation}
where $\rho_+^{(n)}(t) = \text{Im}\ \Pi_{+}(t+i\epsilon) t^{-(n+1)}$. The form factor is an analytic function in the cut complex $t$-plane, so we can apply the standard techniques to derive 
the constraint on the form factor \cite{Ananthanarayan:2011uc}. We can bring Eq.~(\ref{inEqB4map}) to a canonical form by mapping it to the interior of a unit disk using the transformation 
in Eq.~(\ref{Zexp}). In this mapping $z(t_+)=1$ and $z(\infty)=-1$. In terms of $z$, the inequality is
\begin{equation}\label{gBound}
\frac{1}{2\pi}\int_{0}^{2\pi}d\phi \left|g_{+}^{(n)}\left(e^{i\phi}\right)\right|^2 \leq \chi_{+}^{(n)}.
\end{equation}
On one hand the analytic function $g^{(n)}_+(z)$ can be written as 
\begin{equation}\label{FW}
g_{+}^{(n)}\left(z\right)=F_{+}\left(\tilde{t}\left(z\right)\right)\omega_{+}^{(n)}\left(z\right), 
\end{equation}
where $\omega_{+}^{(n)}\left(z\right)$ is called the outer function. The outer function is analytic and has no zeroes in the support domain such that 
$\left|\omega_{+}^{(n)}\right|^2$ is equal to $\rho_+^{(n)}(\tilde{t}(e^{i\phi}))$ times the Jacobian of Eq.~(\ref{Zexp}),
\begin{equation}\label{outer}
\omega_{+}^{(n)}(z)=\left(\frac{1}{4\sqrt{2\pi}}\right)
\frac{\left(t_+ - \tilde{t}(z)\right)\left(\sqrt{t_+ - t_-} + \sqrt{t_+ - \tilde{t}(z)}\right)^{\frac{3}{2}}\left(\sqrt{t_+ - t_0} + \sqrt{t_+ - \tilde{t}(z)}\right)}
{\sqrt[4]{t_+-t_0}\left(\sqrt{t_+} + \sqrt{t_+ - \tilde{t}(z)}\right)^{(n+4)}}
\end{equation}
where we have set $t_0=0$ for our calculation. The outer function and the form factor can be expressed in terms of the variable $z$ using $\tilde{t}(z)=t_+\left(1-\frac{(1-z)^2}{(1+z)^2}\right)$, 
which is the inverse transform of Eq.~(\ref{Zexp}). On the other hand, $g^{(n)}_+(z)$ can be expanded in $z$,
\begin{equation}\label{gExp}
g_{+}^{(n)}\left(z\right)=g_{+,0}^{(n)} + g_{+,1}^{(n)} z + g_{+,2}^{(n)} z^2 + \dots \ .
\end{equation}
This expansion is convergent for $\left|z\right|<1$. It follows from Eq.~(\ref{gBound}) that the coefficients must satisfy the inequality
\begin{equation}\label{gIneq}
\chi_{+}^{(n)} \geq \sum_{j=0}^{\infty}\left(g_{+,j}^{(n)}\right)^2
\end{equation}
The left side of the above inequality is always positive, which leads to a maximum number of $g$-coefficients being non-zero. 
Expanding Eq.~(\ref{FW}) in a Taylor series around $z=0$ and setting it equal to Eq.~(\ref{gExp}) results in each $g$-coefficients being a function of 
$F_+ (0)$, $F_{1,2}$, and meson masses. Substituting $g_{+,j}^{(n)}(F_+ (0), F_{1,2},...)$ into Eq.~(\ref{gIneq}) and solving for $F_+(0)$ leads to a 
bound on the form factor (at $q^2=0$) in terms of $F_1$, $F_2$, and $\chi_{+}^{(n)}$,
\begin{equation}\label{hIneq}
\left|F_+(0)\right| \leq h^{(n)}\left(\chi_{+}^{(n)}, F_1, F_2\right)
\end{equation}
Where the $h^{(n)}$ is the function that results in solving the inequality for $F_+(0)$. The moments $\chi_{+}^{(n)}$ have been calculated in OPE as the sum of the perturbative and non-perturbative contributions. We calculated updated numbers for the moments based on new values for the condensates and masses at a scale of $\mu = 2 \text{ GeV}$. The parameter values at a scale of $\mu = 2 \text{ GeV}$ were calculated by others and can be found in the table below \cite{Gunawardana:2019gep, Khodjamirian:2017zdu, Narison:2011xe, Tanabashi:2018oca}:
\begin{table} 
 \begin{tabular}{|c|c|} 
 \hline\hline
 ~~~Quantity~~~ &\quad Value  \\ [2ex] 
 \hline
 $\alpha_S (2 \text{ GeV})$ &\quad $0.38 \pm 0.03$  \\ 
 \hline
 $m_{c,\text{pole}}$ &\quad $1.67 \pm 0.07 \text{ GeV}$  \\ 
 \hline
 $\left<\bar{u}u\right> (2 \text{ GeV})$ &\quad $(-0.276^{+0.012}_{-0.010} \text{ GeV})^3$  \\ 
 \hline
 $\left<\alpha G^2\right> $ &\quad $(7.0 \pm 1.3)\times 10^{-2} \text{ GeV}^4$ \\ 
 \hline
 $\bar{m}_c (2 \text{ GeV})$ &\quad $1.10 \pm 0.03 \text{ GeV}$  \\ 
 \hline\hline
\end{tabular} 
\caption{Perturbative and non perturbative parameters used in the calculation of moments $\chi_+$.}
\label{table:chiParams}
\end{table}
The perturbative pieces for the heavy to light correlators were calculated for up to two loops in \cite{Chetyrkin:2001je}. Using Eqs.~(34), (35), and the Appendix of \cite{Chetyrkin:2001je}, we can find the updated 
values for $\chi_{+}^{(n)PT}$, for which we include uncertainties. The results are in the Table \ref{table:chiValues}. The non-perturbative piece can be written as
\begin{equation}
\chi_+^{(n)NP}=-\frac{1}{m_{c,\text{pole}}^{2(n+2)}} \left[\bar{m}_c \left<\bar{u}u\right> + \frac{\left<\alpha G^2\right>}{12\pi}\right]
\end{equation}
where $\left<\bar{u}u\right>$ and $\left<\alpha G^2\right>$ are the quark and gluon condensates, respectively. These parameter values have been taken 
from Table \ref{table:chiParams}, and the updated values for the moments can be found in Table \ref{table:chiValues}.

\begin{table} 
 \begin{tabular}{|c|c|c|c|} 
 \hline\hline
 ~~Moment, n~~ &\quad $\chi_+^{(n)NP}\times 10^{-3}$ &\quad $\chi_+^{(n)PT}\times 10^{-3}$ &\quad $\chi_+^{(n)} \times 10^{-3}$ \\ [2ex] 
 \hline
 $1  \left(\text{in GeV}^{-2}\right)$ &\quad $0.98 \pm 0.25$ &\quad $6.37 \pm 0.67$ &\quad $7.35 \pm 0.89$ \\ 
 \hline
 $2  \left(\text{in GeV}^{-4}\right)$ &\quad $0.35 \pm 0.12$ &\quad $0.80 \pm 0.15$ &\quad $1.15 \pm 0.26$ \\ 
 \hline
 $3  \left(\text{in GeV}^{-6}\right)$ &\quad $0.13 \pm 0.05$ &\quad $0.14 \pm 0.04$ &\quad $0.27 \pm 0.09$ \\ 
 \hline\hline
\end{tabular}
\caption{Perturbative and non perturbative parts of the moments $\chi_+^{(n)}$ for $n=1,2,3$.}
\label{table:chiValues}
\end{table}

We have also found $F_{1}$ and $F_{2}$ for our averaged network by fitting the calculated data to a Taylor expansion around $q^2=0$. Only using the data range 
$0 \leq q^2 \leq 0.74 \text{ GeV}^{2}$, we found our two fits to be $F_{1}=(2.062 \pm 0.261) \times 10^{-1} \text{ GeV}^{-1}$ and $F_{2}=(0.869 \pm 0.290) \times 10^{-1} \text{ GeV}^{-2}$,
as shown in Table \ref{table:RatioTable}. Using these two values with $\left|V_{cd} F_+(0)\right| = (14.92 \pm 0.14)\times 10^{-2}$, and plugging it into the inequality we 
obtained with Eq.~(\ref{hIneq}), we can find an upper bound for the form factor at $q^2=0$ for each moment that has been calculated. The results can be found in Table \ref{table:FandVbound}.
The results quoted in the table are consistent with the result $\left|V_{cd}\right| = 0.218 \pm 0.004$ quoted by the Particle Data Group (PDG) \cite{Tanabashi:2018oca}.

\begin{table} 
 \begin{tabular}{|c|c|c|} 
 \hline\hline
 ~~Moment, n~~ &\quad $\left|F_{+}(0)\right|$, upper bound &\quad $\left|V_{cd}\right|$, lower bound \\ [2ex] 
 \hline
 1 & $1.49 \pm 1.13$ &\quad $0.100 \pm 0.077$  \\ 
 \hline
 2 & $2.05 \pm 1.32$ &\quad $0.073 \pm 0.047$  \\ 
 \hline
 3 & $3.25 \pm 1.70$ &\quad $0.046 \pm 0.024$  \\ 
 \hline\hline
\end{tabular} 
\caption{Upper bound for $\left|F_{+}(0)\right|$  and lower bound for $\left|V_{cd}\right|\times 10^{-2}$ calculated for each moment.}
\label{table:FandVbound}
\end{table}
%

\section{Conclusions}

Accurate theoretical description of semileptonic form factor $F_+(q^2)$ are needed for accurate extraction of the CKM matrix elements $V_{cd}$ and for studies of possible 
new physics contributions. While lattice QCD and QCD sum rules' calculations provide model-independent results for various portions of available $q^2$ range, 
extrapolations of $F_+(q^2)$ are often needed to extend the predictions to other values of $q^2$, for which a particular shape of of the $q^2$-dependence is often used.
What systematic uncertainty does choosing a particular function to describe a $q^2$ dependence of the form factors brings to such extrapolation? 
We performed the fit of the available experimental data to an artificial neural net, which was used in a capacity of universal unbiased approximant. We trained a perceptron 
neural net with two hidden layers of one hundred nodes in each. The results of the ANN training and relevant graphs are available at \url{s.wayne.edu/HEPMachineLearning}.
While the simple ANNs employed in this paper do not provide spectacular extrapolation to $q^2=0$, the obtained results, displayed in Table~\ref{table:RatioTable}, can be 
used to test existing models of $q^2$-dependence of the $F_+(q^2)$ form factor. Based on our fits, we conclude that it is possible that more effective poles need to be taken 
into account if model-inspired form factors are used for parameterizations of future experimental data. Finally, we used the resulting ANN fit to improve unitarity constraints on the 
form factor, which allowed for model-independent bounds on $V_{cd}$. 

This work was supported in part by the U.S. Department of Energy under contract de-sc0007983. We thank Gil Paz for reading the manuscript and helpful comments.
AAP thanks Roy Briere for useful and oftentimes sobering conversations. AAP thanks the Institute for Nuclear Theory at the University of Washington for its kind hospitality and 
stimulating research environment. This research was also supported in part by the INT's U.S. Department of Energy grant No. DE-FG02- 00ER41132.



\begin{thebibliography}{99}

\bibitem{Artuso:2008vf} 
  M.~Artuso, B.~Meadows and A.~A.~Petrov,
  Ann.\ Rev.\ Nucl.\ Part.\ Sci.\  {\bf 58}, 249 (2008)

\bibitem{Ablikim:2018evp} 
  M.~Ablikim {\it et al.} [BESIII Collaboration],
  Phys.\ Rev.\ Lett.\  {\bf 122}, no. 1, 011804 (2019)

\bibitem{Ablikim:2018frk} 
  M.~Ablikim {\it et al.} [BESIII Collaboration],
  Phys.\ Rev.\ Lett.\  {\bf 121}, no. 17, 171803 (2018)
  
\bibitem{Yuan:2019zfo} 
  C.~Z.~Yuan and S.~L.~Olsen,
  Nature Rev.\ Phys.\  {\bf 1}, no. 8, 480 (2019).

\bibitem{Riggio:2017zwh} 
  L.~Riggio, G.~Salerno and S.~Simula,
  Eur.\ Phys.\ J.\ C {\bf 78}, no. 6, 501 (2018)

\bibitem{Amhis:2019ckw} 
  Y.~S.~Amhis {\it et al.} [HFLAV Collaboration],
  arXiv:1909.12524 [hep-ex].

\bibitem{Aoki:2019cca} 
  S.~Aoki {\it et al.} [Flavour Lattice Averaging Group],
  arXiv:1902.08191 [hep-lat].

\bibitem{Khodjamirian:2009ys} 
  A.~Khodjamirian, C.~Klein, T.~Mannel and N.~Offen,
  Phys.\ Rev.\ D {\bf 80}, 114005 (2009)


\bibitem{Boyd:1994tt} 
  C.~G.~Boyd, B.~Grinstein and R.~F.~Lebed,
  Phys.\ Rev.\ Lett.\  {\bf 74}, 4603 (1995)

\bibitem{Becher:2005bg} 
  T.~Becher and R.~J.~Hill,
  Phys.\ Lett.\ B {\bf 633}, 61 (2006)

\bibitem{Ananthanarayan:2011uc} 
  B.~Ananthanarayan, I.~Caprini and I.~Sentitemsu Imsong,
  Eur.\ Phys.\ J.\ A {\bf 47}, 147 (2011)

\bibitem{Grinstein:2015wqa} 
  B.~Grinstein and R.~F.~Lebed,
  Phys.\ Rev.\ D {\bf 92}, no. 11, 116001 (2015)
  
\bibitem{DescotesGenon:2008hh} 
  S.~Descotes-Genon and A.~Le Yaouanc,
  J.\ Phys.\ G {\bf 35}, 115005 (2008)


\bibitem{Faustov:2019mqr} 
  See, e.g., R.~N.~Faustov, V.~O.~Galkin and X.~W.~Kang,
  arXiv:1911.08209 [hep-ph].

\bibitem{Becirevic:1999kt} 
  D.~Becirevic and A.~B.~Kaidalov,
  Phys.\ Lett.\ B {\bf 478}, 417 (2000)

\bibitem{Ball:2004ye} 
  P.~Ball and R.~Zwicky,
  Phys.\ Rev.\ D {\bf 71}, 014015 (2005)

\bibitem{Su:2010my}
F.~Su and Y.~Yang,
Int. J. Mod. Phys. A \textbf{26}, 3185-3199 (2011)



\bibitem{Kolmogorov57}
A.~N.~Kolmogorov,
Dokl. Akad. Nauk. SSSR, 114 (1957), 953-956

\bibitem{NNEstimator} 
  K.~Hornik, M.~Stinchcombe, and H.~White,
  Neural Networks {\bf 2}, no. 5, 359-366 (1989)

\bibitem{NNEstimator2} 
  G.~Cybenko
  Math. Control Signals Systems (1989) {\bf 2} 303-314
   
\bibitem{Forte:2002fg} 
  S.~Forte, L.~Garrido, J.~I.~Latorre and A.~Piccione,
  JHEP {\bf 0205}, 062 (2002)
 
\bibitem{Ball:2014uwa} 
  R.~D.~Ball {\it et al.} [NNPDF Collaboration],
  JHEP {\bf 1504}, 040 (2015)

\bibitem{Rojo:2006nn} 
  J.~C.~Rojo,
  ``The Neural network approach to parton distribution functions,''
  hep-ph/0607122.

\bibitem{Graczyk:2010gw} 
  K.~M.~Graczyk, P.~Plonski and R.~Sulej,
  JHEP {\bf 1009}, 053 (2010)

\bibitem{Alvarez-Ruso:2018rdx} 
  L.~Alvarez-Ruso, K.~M.~Graczyk and E.~Saul-Sala,
  Phys.\ Rev.\ C {\bf 99}, no. 2, 025204 (2019)
 
\bibitem{Carleo:2019ptp} 
  G.~Carleo, I.~Cirac, K.~Cranmer, L.~Daudet, M.~Schuld, N.~Tishby, L.~Vogt-Maranto and L.~Zdeborova,
  Rev.\ Mod.\ Phys.\  {\bf 91}, no. 4, 045002 (2019)

\bibitem{Nocedal:2000sp} 
  J.~Nocedal and S.~J.~Wright,
  doi:10.1007/b98874
  
\bibitem{CGmethod} 
  M.~F.~Moller, 
  Neural Networks {\bf 6}, no. 4, 525-533 (1993)

\bibitem{Demuth:2014} 
  Demuth, Howard B. and Beale, Mark H. and De Jess, Orlando and Hagan, Martin T.,
``Neural Network Design,''
ISBN 0971732116, 9780971732117

\bibitem{Ablikim:2015ixa} 
  M.~Ablikim {\it et al.} [BESIII Collaboration],
  Phys.\ Rev.\ D {\bf 92}, no. 7, 072012 (2015)

\bibitem{patro2015normalization} 
  S.~G.~Krishna and K.~K.~Sahu,
  arXiv:1503.06462 [cs.OH].
  
\bibitem{Paz:2019wfq} 
  G.~Paz,
  arXiv:1909.08108 [hep-ph].
  
\bibitem{Gunawardana:2019gep} 
  A.~Gunawardana and G.~Paz,
  JHEP {\bf 1911}, 141 (2019)

\bibitem{Khodjamirian:2017zdu} 
  A.~Khodjamirian and A.~A.~Petrov,
  Phys.\ Lett.\ B {\bf 774}, 235 (2017)

\bibitem{Narison:2011xe} 
  S.~Narison,
  Phys.\ Lett.\ B {\bf 706}, 412 (2012)
  
\bibitem{Tanabashi:2018oca} 
  M.~Tanabashi {\it et al.} [Particle Data Group],
  Phys.\ Rev.\ D {\bf 98}, no. 3, 030001 (2018).
  
\bibitem{Chetyrkin:2001je} 
  K.~G.~Chetyrkin and M.~Steinhauser,
  Eur.\ Phys.\ J.\ C {\bf 21}, 319 (2001)
    
\end{thebibliography}
\end{document}